# Movement Synchronization in Complex and Dynamic Team Coordination

**JUN ICHIKAWA[1], MIZUKI YOKOYAMA[2], SOMA ISHIDA[2], and GENKI ICHINOSE[2]**

[1]Faculty of Informatics, Shizuoka University, Hamamatsu 432-8011 JAPAN

[2]Faculty of Engineering, Shizuoka University, Hamamatsu 432-8561 JAPAN

Corresponding author: Jun Ichikawa (e-mail: j-ichikawa@inf.shizuoka.ac.jp).

This study was supported by JSPS KAKENHI Grant Number 24K20562 and Daiichi-Sankyo "Habataku" Support Program for the Next Generation of Researchers.

**ABSTRACT** Mutual anticipation of teammates' actions enables efficient coordination that achieves a common goal and high performance. In team sports involving direct competition, such implicit and nonverbal interactions within short periods are required. If players begin moving after observing their teammates, gaps may emerge, allowing opponents to interfere. When mutual anticipation functions appropriately, players' interactions are smooth without gaps, and their movements are expected to become synchronized. Synchronization represents a temporally stable structure whose mechanisms have been examined. However, few studies have investigated synchronization in complex and dynamic real-world coordination involving heterogeneous roles and evolving interactions within short periods. The examinations have been essentially limited to determining whether there are statistical differences from a baseline. In our approach, we utilized team sports and introduced a statistical method to provide detailed probabilistic evaluations of these differences. The purpose of this study was to extract the temporal components of movement using 3-on-3 basketball. In a field experiment investigating offensive coordination, we calculated the relative phases in which players approached or moved away from their teammates during mini-games. The frequency distributions were estimated using Bayesian inference and compared before and after receiving advice from the coach. The results showed that the posterior probabilities that the frequencies in synchronization-related bins were higher after the advice than before the advice were 70% or higher. As a pilot study, these findings may provide a first step toward understanding coordination processes related to mutual anticipation based on a common ground of team strategy.

## I. INTRODUCTION

Humans often achieve a common goal and high performance through cooperation (e.g., [1]–[3]). For such team coordination, mutual anticipation of actions enables implicit and nonverbal interactions [4], [5]. Efficient interactions within short periods are required in team sports involving direct competition. If players begin moving after observing their teammates, gaps may emerge, allowing opponents to interfere. Basic research in human movement science and perceptual psychology suggests a relationship between expertise and anticipation. Experienced players anticipate subsequent actions more quickly and accurately than novices [6]–[8]. If mutual anticipation functions appropriately, players' interactions are smooth without gaps, and their movements are expected to become synchronized. Synchronization generally refers to a temporally stable structure of interactions involving two or more individuals [9], [10]. This cognitive science study investigated synchronization and discussed its potential relationship with mutual anticipation.

Previous studies in psychology and cognitive science suggested a relationship between mutual anticipation and movement synchronization and discussed a common ground of activities underlying these processes. A common ground enables team members to simulate and anticipate behaviors appropriately [2], [11], [12]. A representative concept is a shared mental model, which refers to structured and shared knowledge and beliefs [5], [13], [14]. Team cognition has also been treated as information processing, in which the knowledge distributed among members is integrated and mutually utilized [15], [16]. As explained above, these cognitive mechanisms enable implicit interactions without verbal communication [4], [5]. Therefore, synchronization may serve as a potential marker for discussing mutual anticipation. Meanwhile, synchronization has been observed in humans, other organisms, and non-living systems such as clapping, walking, firefly flashing, and metronome pendulums (e.g., [17]–[20]). These characteristics can be constructively explained using mathematical models such as the Kuramoto model [21]. In-phase and anti-phase synchronizations are representative examples of temporally stable structures. When movement phase is represented as an angle on a circle, in-phase synchronization is defined as a relative phase of 0 degrees between two individuals, indicating that their rhythms are perfectly aligned. Anti-phase synchronization is defined by a relative phase of −180 or 180 degrees, indicating a difference of exactly half a cycle [22]. Synchronization studies suggest adaptive functions for trust-related prosocial behaviors, empathy, and problem solving in humans [23]–[26].

Some studies have investigated movement synchronization in sports, which may appear difficult to identify at first glance. Synchronization with teammates and opponents has been observed in a three-versus-one ball possession task in soccer [27], a competitive play-tag game [28], a kendo match [29], and a breakdance battle [9], [10]. Synchronization with opponents may reflect mutual adjustment, which is established as a game. Thus, synchronization supports high performance. However, few studies have investigated synchronization in more complex and dynamic real-world coordination involving heterogeneous roles and evolving interactions within short periods. For example, in a three-versus-one ball possession task in soccer, a team is arranged in a triangular formation without heterogeneous roles such as defender, midfielder, and forward. Recently, synchronization has been examined using big data analysis of professional soccer league matches [30]. However, evaluation methods have been essentially limited to determining whether there are statistical differences from a baseline using statistical hypothesis tests.

In our approach, we utilized team sports and introduced a statistical method to provide more detailed probabilistic evaluations of these differences in synchronization. The purpose of this study was to extract the temporal components of movement using 3-on-3 basketball. This study used a dataset from a field experiment conducted at a single university to investigate offensive coordination. We calculated the relative phases in which players approached or moved away from their teammates during mini-games. These frequency distributions were estimated using Bayesian inference and compared before and after receiving advice from the coach. This pilot study is expected to contribute to quantitatively understanding coordination processes.

## II. METHODS

### A. DATASET

We reused the dataset recorded by Ichikawa et al. [31] (see the original description under CC BY ver. 4.0). In this previous study, a field experiment using 3-on-3 basketball was conducted to examine the crucial role of intervention decision and adjustment in team coordination (see Section II-B) [1]. The dataset consists of two-dimensional time-series positions of each player obtained from a bird's-eye view video using a single camera to capture all players and the entire court (950 cm × 1505 cm) for a total of 42 games (trials).

Players recorded at 20 fps were tracked using ByteTrack [32] based on YOLOX. False detections and undetected positives were manually corrected using the labeling platform Labelbox [2]. The two-dimensional positions of each player, originally recorded in pixel units, were converted to centimeter units through a projective transformation matrix based on the real-world coordinates of 12 reference points, including the four corners of the court. The mean absolute measurement error was 3.165 cm × 2.102 cm, comparable to the accuracy (2.3 cm × 4.6 cm) reported in a previous study [27] that examined movement synchronization in a three-versus-one ball possession task in soccer.

### B. OVERVIEW OF FIELD EXPERIMENT AND RESULTS

We provide an overview of the field experiment conducted by Ichikawa et al. [31] (see the original description under CC BY ver. 4.0). Six female students from a university basketball club, competing in the third division of the Tokai Area League in Japan, participated in the experiment. They were divided into offensive and defensive teams and engaged in mini-games. Figure 1 illustrates a diagram of the mini-game used to examine offensive coordination. The offensive team's win criterion was to create an open space for an unmarked shot within a 15-s time limit by having an offensive player draw a defensive player. The defensive team was required to prevent the offensive team from meeting this criterion. The offensive team was required to perform the fundamental coordination common in 5-on-5 basketball, starting from the initial position, as shown in Figure 1(a). The role-sharing structure was as follows: Offensive #1 primarily attacks by dribbling toward the goal, Offensive #2 cooperates with Offensive #1 to create a shot chance, and Offensive #3 adjusts the two-man game to create favorable situations. This player moderately intervenes with the teammates according to the situation.

Figure 1 also shows some examples of coordination involving Offensive #3, such as (b) "hand-off," (c) "pick and roll," and (d) running to a corner area to attempt a three-point shot. "Hand-off" refers to approaching an on-ball player and receiving a pass directly. "Pick and roll" refers to sequential interactions in which a player sets a screen (pick) against a defensive player, and a teammate uses the pick to evade the defense and dribbles toward the goal. In addition, Offensive #3 is also required (e) to stay without disrupting a teammate's play space. As a pretest, three sessions comprising seven game trials per session were conducted for a total of 21 trials. After the pretest, a guest coach provided advice to the offensive team, as explained above. A posttest was conducted in a manner similar to the pretest. The results showed that offensive performance in each session, as indicated by the win rate, remained at the chance level (0.571) in the pretest, whereas it improved to a maximum of 0.857 in the posttest. Furthermore, using the tracking data, the distance between Offensive #3, who played the crucial role of intervention decision and adjustment, and each teammate or opponent was calculated at each time frame. The frequencies in the bins with relatively large distances tended to be significantly higher in the posttest condition than those in the pretest condition. These characteristics suggest that Offensive #3 maintained an appropriate distance from the teammates and opponents to implement her role. Based on the analysis and video recordings, the offensive team appeared to exhibit the coordination pattern (d) shown in Figure 1 (Figure 2).

[1] This previous study was approved by the ethics and safety committee of the institutions with which the first author and participants were affiliated. Written informed consent regarding video recording and data collection was obtained from all the participants.

[2] https://labelbox.com/

The offensive team may have established the specific coordination pattern in the posttest. Thus, their movements are predicted to become synchronized. The duration of the game varied across the trials. The average durations in the pretest and posttest conditions were 8.812 s ($SD = 3.330$) and 9.026 s ($SD = 3.141$), respectively. Each trial had a 15-s time limit and ended when an offensive player attempted a shot, the ball went out of bounds, or a defensive player disrupted offensive coordination. The analysis interval was defined as the period from the start to end of each trial. In this analysis, the definition was applied consistently across all the trials. In addition, considering that the videos were at 20 fps and that this study utilized team sports in which interactions among members evolved within short periods, no filtering or smoothing was applied, as in the previous study [31].

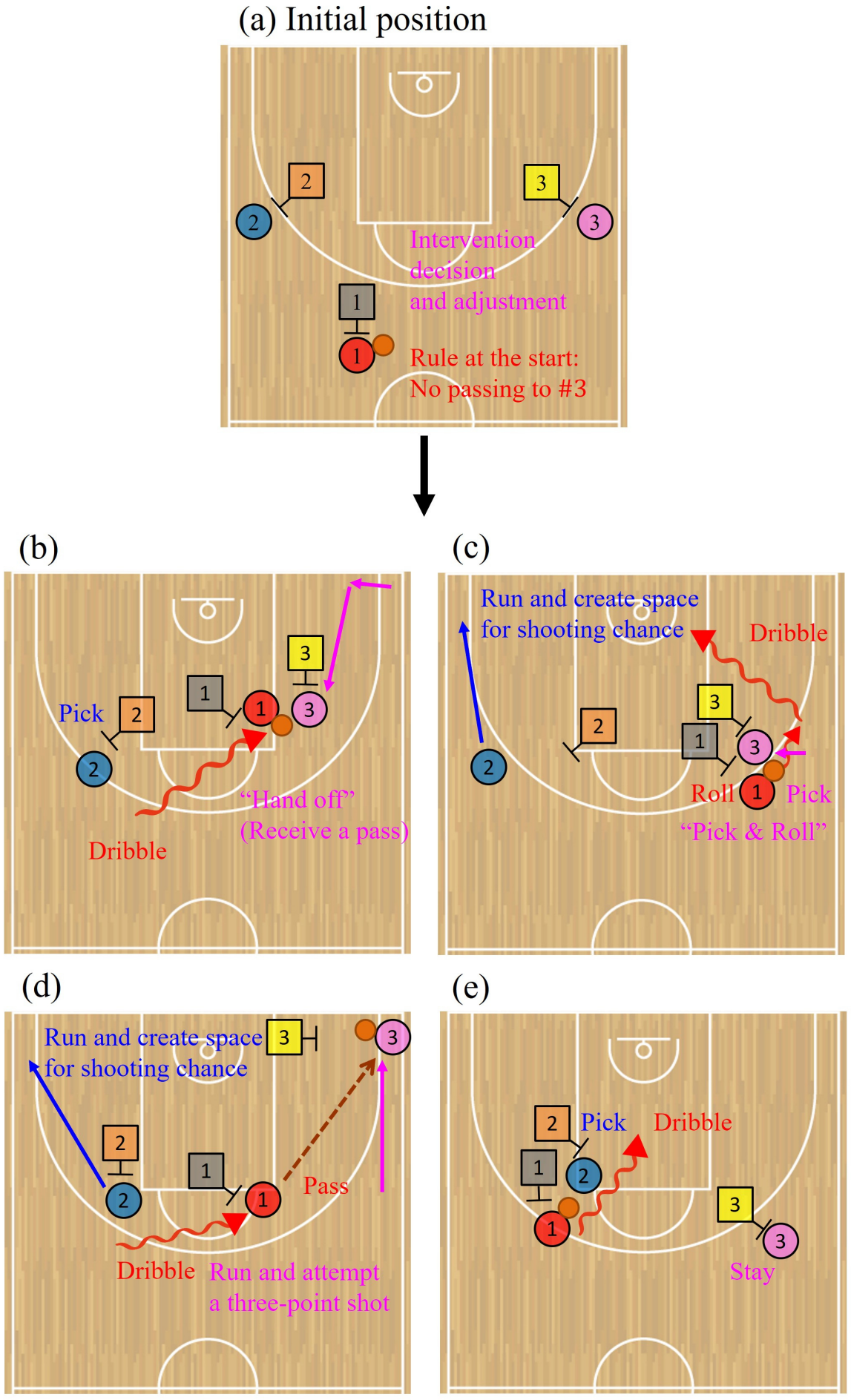

**FIGURE 1. Diagram of the mini-game in the field experiment conducted by Ichikawa et al. [31]. Circles and squares represent offensive and defensive players, respectively. Each color indicates the bib worn to identify each individual. (a) shows an example of the initial position. (b)–(e) show offensive coordination related to the crucial role of intervention decision and adjustment (Offensive #3). This figure is modified from the study [33] under CC BY ver. 4.0.**

### C. ANALYSIS

#### 1) RELATIVE PHASE

The relative phase was calculated based on previous studies [9], [10], [28], [29]. Figure 3 illustrates a diagram of the analysis procedures for Offensive #1 and #3. At each time frame $t$, we calculated the movement vectors of the players $\boldsymbol{M}^{O1}_{t-1\rightarrow t+1}$ and $\boldsymbol{M}^{O3}_{t-1\rightarrow t+1}$. The inter-player distance vectors $\boldsymbol{D}_t^{O1O3}$ and $\boldsymbol{D}_t^{O3O1}$ were also computed. $\boldsymbol{D}_t^{O1O3}$ represents the vector from Offensive #1 to Offensive #3, and $\boldsymbol{D}_t^{O3O1}$ denotes the vector in the opposite direction. Using these vectors, each player's vector is projected in the direction of the distance vector toward the teammate based on Eq. (1). The temporal components $V_t^{O1}$ and $V_t^{O3}$, indicating movements toward or away from the teammate, are extracted.

$$
\begin{aligned}
V_t^{O1} &= \frac{\boldsymbol{M}^{O1}_{t-1\rightarrow t+1} \cdot \boldsymbol{D}_t^{O1O3}}{\boldsymbol{D}_t^{O1O3} \cdot \boldsymbol{D}_t^{O1O3}} |\boldsymbol{D}_t^{O1O3}|, \\
V_t^{O3} &= \frac{\boldsymbol{M}^{O3}_{t-1\rightarrow t+1} \cdot \boldsymbol{D}_t^{O3O1}}{\boldsymbol{D}_t^{O3O1} \cdot \boldsymbol{D}_t^{O3O1}} |\boldsymbol{D}_t^{O3O1}|.
\end{aligned} \tag{1}
$$

Calculating inter-individual distances is a fundamental analysis for capturing the dynamics of group movement and has been widely used not only in team sports but also in collective motion and children's play [35]–[37]. In team sports, previous studies suggested that movements that create and maintain appropriate distances and spaces between teammates and opponents are important for offensive coordination [38]–[40]. In our previous study [31], a tendency toward differences between the pretest and posttest conditions was also observed in the distances between the offensive player in the crucial role of intervention decision and adjustment and the other players. It may be related to the specific pattern of offensive coordination (Figure 2). Based on these findings, we analyzed movement synchronization by focusing on the distances between the offensive players. Meanwhile, examining synchronization from various perspectives, such as lateral or tangential movement components (i.e., components perpendicular to the inter-player distance vector), should be addressed in future studies.

Furthermore, the time series of $V^{O1}$ and $V^{O3}$ are standardized, and the movement phases at the current time $t$ $M_t^{O1} e^{i\phi_t^{O1}}$

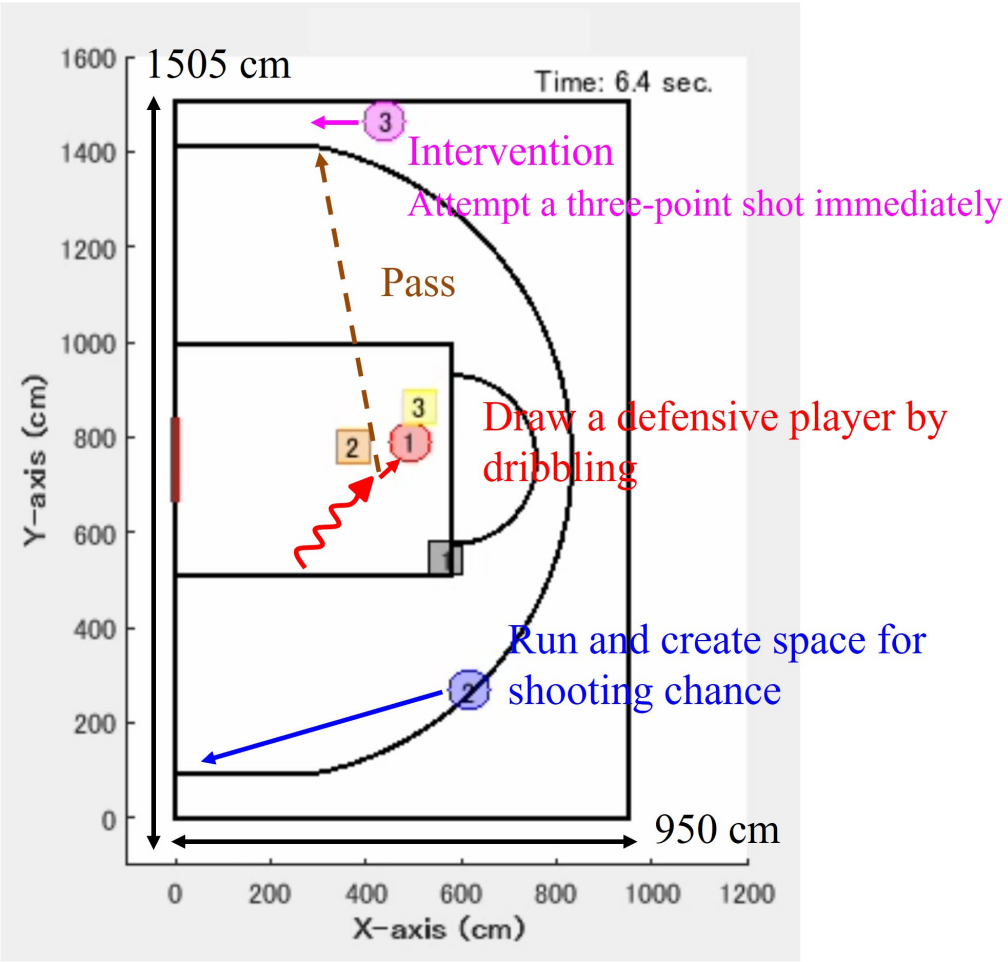

**FIGURE 2. Example of offensive coordination that the team appeared to exhibit in the posttest condition conducted by Ichikawa et al. [31]. This corresponds to the pattern, as shown in Figure 1(d). The figure is created using the dataset with reference to the study [34].**

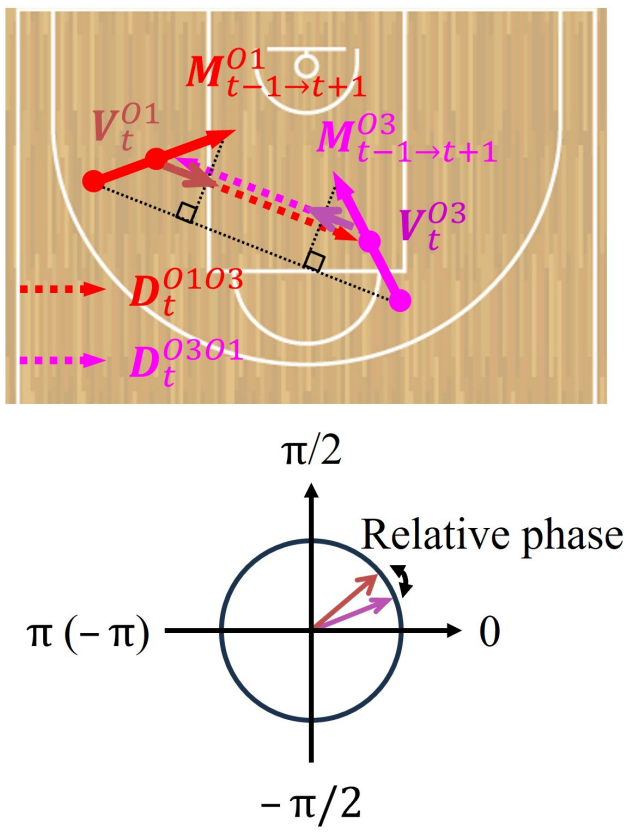

**FIGURE 3. Diagram of the analysis procedures between the pair of Offensive #1 and #3. The symbols correspond to Eq. (1).**

and $M_t^{O3} e^{i\phi_t^{O3}}$ are extracted using the Hilbert transform in Eq. (2).

$$\begin{aligned} V_t^{O1} &= P_t^{O1} + i\hat{P}_t^{O1} = M_t^{O1} e^{i\phi_t^{O1}}, \\ V_t^{O3} &= P_t^{O3} + i\hat{P}_t^{O3} = M_t^{O3} e^{i\phi_t^{O3}}. \end{aligned} \tag{2}$$

The relative phase $\Delta\phi_t^{O1O3}$ is then calculated by (3).

$$\Delta\phi_t^{O1O3} = \text{Arg}\left(e^{i(\phi_t^{O1} - \phi_t^{O3})}\right). \tag{3}$$

$\Delta\phi_t^{O1O3}$ ranges from $-\pi$ (−180 degrees) to $\pi$ (180 degrees). A relative phase of 0 degrees indicates in-phase synchronization, whereas a relative phase of −180 or 180 degrees denotes anti-phase synchronization. When Offensive #1 and #3 simultaneously approach or move away from each other, the relative phase is 0 degrees. In contrast, when one player approaches a teammate while the teammate simultaneously moves away, the relative phase is −180 or 180 degrees, indicating a difference of exactly half a cycle. The relative phase was calculated for each time frame for all the offensive pairs in each game trial: Offensive #1 and #2, Offensive #1 and #3, and Offensive #2 and #3. We created histograms with 20-degree bins to represent the frequency distributions in the pretest and posttest conditions. For reference, we also created frequency distributions using different bins (10- and 30-degree bins) for the sensitivity analysis.

As shown in Eq. (1), the movement vector at time $t$ was calculated from the difference in position between $t-1$ and $t+1$. Thus, for the first and last time frames, the movement vectors calculated from the adjacent time frames were assigned as boundary treatment. Subsequently, no specific boundary treatment was applied for the Hilbert transform, and no exceptional treatment was applied to periods of low-speed and stationary movements. This analysis procedure itself was based on previous studies that examined movement synchronization using sports, as in this study [9], [10], [28], [29]. For boundary treatment, the padding methods and padding length may influence the results. Careful consideration is required to determine appropriate procedures for the dynamic characteristics of team sports data, in which interactions among players can evolve within short periods. The mini-game began with the offensive player (Offensive #1) possessing the ball, and the team attempted to create a shooting opportunity through coordination. Although periods of low-speed and stationary movements were limited, these actions could include strategically meaningful play, such as not intervening to avoid interfering with a teammate, as shown in Figure 1(e). As the Hilbert transform calculates the instantaneous phase based on the entire time-series, arbitrarily excluding only low-speed and stationary time frames can influence the results. Nevertheless, the boundary effects associated with the Hilbert transform and exceptional handling of low-speed and stationary movements should be further examined in future studies.

### 2) BAYESIAN INFERENCE

For the pretest and posttest conditions, the average frequency distributions of the relative phases across the 21 game trials were computed. Statistical hypothesis tests, such as $t$-tests and ANOVA, have often been used to compare frequencies. For example, in normalized frequency distributions, the averages are repeatedly compared between experimental conditions or bins (e.g., [9], [10], [29]). However, this evaluation method carries the risk of a Type I error, in which a true null hypothesis of no difference is incorrectly rejected, even when effect sizes are calculated. For example, Shimizu and Okada examined movement synchronization between performers in breakdance battles [9], [10]. Using the frequencies in the bins around −180 and 180 degrees of relative phases as a baseline, representing anti-phase synchronization, they compared the frequencies with those in other bins.

Furthermore, the interpretation of the differences is es-sentially limited to determining whether there are statistical differences from the baseline. Therefore, we estimated the frequency distribution for each condition from the observed data using Bayesian inference and calculated the certainty of the direction of the difference from the pretest condition in each bin. This enabled a more detailed probabilistic evaluation. Bayesian inference allows the distribution of observed data to be flexibly modeled unlike a $t$-test and ANOVA, which assume a normal distribution [41], [42]. In contrast, the effectiveness of other approaches such as machine learning for datasets with a limited sample size (number of trials in this study), has not been sufficiently examined. Thus, the optimal method for evaluating the differences in synchronization between the conditions should be further investigated in future studies.

In this analysis, the vector of the frequency distribution with 20-degree bins in trial $n$ ($1 \le n \le 21$) is defined as $\boldsymbol{y}_n = (y_n^1, ..., y_n^{18})$. $\boldsymbol{y}_n$ represents the observed data. To incorporate information regarding the number of observations such as the number of time frames into the estimation, the frequencies were not normalized at this stage. The true distribution in trial $n$ is denoted by $\boldsymbol{p}_n = (p_n^1, ..., p_n^{18})$. $\boldsymbol{p}_n$ is a latent vector indicating the probability of each observation being assigned to a bin. Subsequently, $\boldsymbol{y}_n$ is given by Eq. (4).

$$\boldsymbol{y}_n \sim \text{Multinomial}(N_n, \boldsymbol{p}_n). \tag{4}$$

$N_n$ denotes the total number of observations in trial $n$, given by $\sum_{k=1}^{18} y_n^k$. In Bayesian inference, the likelihood represents the probability of the observed data given the parameters, whereas the prior distribution represents the assumptions regarding these parameters before observing the data. By combining these components, the posterior distribution of the parameters is updated after observing the data, as shown in Eq. (5).

$$p(\boldsymbol{p}_n \mid \boldsymbol{y}_n) = \frac{p(\boldsymbol{y}_n \mid \boldsymbol{p}_n)p(\boldsymbol{p}_n)}{p(\boldsymbol{y}_n)}. \tag{5}$$

$p(\boldsymbol{p}_n)$ is the prior distribution and $p(\boldsymbol{y}_n \mid \boldsymbol{p}_n)$ is the likelihood. $p(\boldsymbol{y}_n)$ is the marginal likelihood used to normalize the posterior distribution. $p(\boldsymbol{p}_n \mid \boldsymbol{y}_n)$ represents the posterior distribution.

Furthermore, $\boldsymbol{m}_c = (m_c^1, ..., m_c^{18})$ is introduced as a probability vector indicating the average frequency distribution across the trials under condition $c$. The Dirichlet distribution shown in Eq. (6) is also specified as the prior distribution, where $k$ denotes each bin ($1 \le k \le 18$).

$$\boldsymbol{m}_c \sim \text{Dirichlet}(1^k). \tag{6}$$

$\boldsymbol{p}_n$ is then assumed to follow the Dirichlet distribution with average $\boldsymbol{m}_c$, as shown in Eqs. (7) and (8).

$$\boldsymbol{p}_n \sim \text{Dirichlet}(\phi_c \boldsymbol{m}_c), \tag{7}$$

$$\phi_c \sim \text{Lognormal}(2, 1). \tag{8}$$

The concentration parameter $\phi_c$ is introduced to explain the variability across the trials. Here, a log-normal prior distribution with a mean of 2 and a standard deviation of 1 on the log scale is set. The larger this value, the more concentrated $\boldsymbol{p}_n$ becomes around $\boldsymbol{m}_c$, indicating smaller variability across the trials.

Based on the model described above, the unobserved parameters $\boldsymbol{m}_c$ and $\phi_c$ were estimated. This analysis was conducted using the standard Markov chain Monte Carlo (MCMC) method. Four chains were run with 2000 iterations per chain. The first 1000 iterations of each chain were discarded as the burn-in period (warm-up), and the remaining samples were used for the estimation. An upper limit of 200 was set for $\phi_c$. This value was chosen to prevent excessively large values from destabilizing MCMC sampling, while allowing sufficient variability in the frequency distributions across the trials.

### 3) COMPARISON BETWEEN EXPERIMENTAL CONDITIONS AND HYPOTHESIS

The probability vectors for the average frequency distributions in the pretest and posttest conditions, $\boldsymbol{m}_{pre}$ and $\boldsymbol{m}_{post}$, respectively, were estimated using Bayesian inference. To examine the difference between the pretest and posttest conditions in each bin, $p(\delta^k > 0 \mid data)$ was calculated, where $\delta^k$ is defined as $m_{post}^k - m_{pre}^k$. $data$ refer to all the observations, including the conditions. It represents the certainty of the direction of the difference. A larger value of $p(\delta^k > 0 \mid data)$ indicates a higher posterior probability that the frequency in the posttest condition exceeds that in the pretest condition. Ichikawa et al. [31] suggest that the offensive team may have exhibited the specific coordination pattern in the posttest condition (Figure 2). The team may engage in smoother interactions than in the pretest condition, and their movements are expected to become more synchronized. Therefore, the following hypothesis was tested:

- In the bins around relative phases of 0 degrees, −180 degrees, or 180 degrees, $p(\delta^k > 0 \mid data)$ is higher.

## III. RESULTS

Figure 4 shows the average frequency distributions of the relative phases across the game trials in the pretest and posttest conditions. The horizontal axis represents the 20-degree bins, and the vertical axis represents the normalized frequency. For all the pairs of offensive players, the distributions appeared to differ between the conditions. The tendencies were also observed in the distributions using 10- and 30-degree bins for the sensitivity analysis (see the Supplementary Material).

Next, we report the results of Bayesian inference. The bulk effective sample sizes (ESS) exceeded 5,000 for all the parameters ($\boldsymbol{m}_{pre}$, $\boldsymbol{m}_{post}$, $\boldsymbol{m}_{post} - \boldsymbol{m}_{pre}$, $\phi_{pre}$, and $\phi_{post}$), confirming that sufficient effective sample sizes were obtained. All $\hat{R}$ values used to assess MCMC convergence were below 1.10, indicating appropriate convergence (see the Supplementary Material). Figure 5 shows the frequency distributions of $\boldsymbol{m}_{pre}$ and $\boldsymbol{m}_{post}$. Shaded bands represent 90% credible intervals. These results captured the observed distribution shape under each condition and the relative magnitudes between the conditions. Table 1 presents the estimated values of $\boldsymbol{m}_{post} - \boldsymbol{m}_{pre}$.

Considering the readability of the manuscript, we have summarized the results directly related to the hypothesis testing. The others for $\boldsymbol{m}_{pre}$, $\boldsymbol{m}_{post}$, $\phi_{pre}$, and $\phi_{post}$ are provided in the Supplementary Material.

In addition, Figure 6 shows the posterior probabilities that the frequencies in the posttest condition exceeded those in the pretest condition. The horizontal axis represents the bin, and the vertical axis represents $p(\delta^k > 0 \mid data)$. The dotted line parallel to the horizontal axis indicates the chance level. For Offensive #1 and #2, $p(\delta^k > 0 \mid data)$ was 0.70 or higher in the bins around −180 degrees and 180 degrees as shown by "–", reaching a maximum value above 0.90. This suggests a tendency toward increased anti-phase synchronization in the posttest condition. Furthermore, for Offensive #1 and #3, $p(\delta^k > 0 \mid data)$ was 0.70 or higher in the bins around 0 degrees as shown by "+", reaching a maximum value above 0.90. This suggests a tendency toward increased in-phase synchronization in the posttest condition. In contrast, Offensive #2 and #3 did not show such a characteristic compared to the other pairs. However, $p(\delta^k > 0 \mid data)$ reached nearly 0.95 in the bin around 20 degrees, as shown by "+". It suggests a tentative tendency toward increased in-phase synchronization in the posttest condition.

## IV. DISCUSSION

The results provided suggestive support for our hypothesis. The offensive movements imply a tendency toward increased synchronization in the posttest condition compared to the pretest condition. For Offensive #1 and #2, the posterior probabilities that the frequencies in the posttest condition exceeded those in the pretest condition were 0.70 or higher in the bins around −180 and 180 degrees, reaching a maximum value above 0.90. This suggests a tendency toward increased anti-phase synchronization. For Offensive #1 and #3, the posterior probabilities were 0.70 or higher in the bins around 0 degrees, reaching a maximum value above 0.90. This suggests a tendency toward increased in-phase synchronization. Although Offensive #2 and #3 did not show a pattern like that of the other pairs, the results suggest a tentative tendency toward increased in-phase synchronization.

In team sports in which players are required to interact within short periods to prevent opponents from interfering, moving after observing their teammates' movements may delay movement initiation and affect the relative phase. Ichikawa et al. [31] imply that the team may have exhibited the coordination pattern in which Offensive #3 in the role of intervention decision and adjustment ran to a corner area to attempt a three-point shot in the posttest condition (Figure 2). The offensive players were required to share the heterogeneous roles (see Section II-B). Although each player moved differently, their movements suggest a tendency toward increased synchronization. This may be related to mutual anticipation based on a common ground regarding the specific coordination. However, this is only a possible interpretation in terms of cognitive science. Synchronization does not directly measure the degree of mutual anticipation. Alternative explanations cannot be completely excluded, such as simple familiarization with the repeated mini-games, spatial constraints of the court setting, or the possibility that the players simply reproduced the specific coordination pattern after receiving advice from the coach.

In the posttest condition, the movements of the pairs involving Offensive #1 suggest a tendency toward increased synchronization. Offensive #1 primarily attacked by dribbling toward the goal (see Section II-B). Offensive #1 played a central role in shooting, and it may be related to the characteristics, as shown in Figure 6. Furthermore, the movement for Offensive #1 and #2 suggests a tendency toward increased anti-phase synchronization. The relative phase was −180 or 180 degrees when one player approached the teammate while the teammate simultaneously moved away. It may be related to meaningful space structures. In team sports, creating and controlling effective play space are crucial for high performance [38]–[40]. Therefore, in future studies, we should comprehensively investigate the relationships among synchronization, play events such as passing and dribbling, space structures created through player positioning, and team strategies. For example, this research may be extended to machine-learning methods that predict the success or failure of applying team strategies and the coordination from spatiotemporal data by integrating play events and player positioning, such as the approach of Fujii et al. [43]. The relative phase can then be included as an explanatory variable. Our study should be positioned as a pilot study and a first step toward examining these relationships. It is important for us to conduct continuous measurements and analysis across competitive levels to examine the generalizability and validity of our findings.

This study examined movement synchronization in complex and dynamic team coordination. 3-on-3 basketball involves heterogeneous roles, and the interactions evolve within short periods. We calculated the relative phases and discussed the certainty of the direction of the difference from the baseline condition using Bayesian inference. Our methods can be applied to a wide range of team sports, provided that (1) tracking data containing the two-dimensional time-series positions of each player and (2) a baseline condition are available. Further development is expected to clarify the generalizability of synchronization in coordination by examining various sports. Compared with experimental tasks such as tapping in response to stimuli, team sports make it difficult to subjectively evaluate synchronization. As an application to the sports field, providing detailed feedback on synchronization may provide useful information for advice and practical design. However, as explained above, it should be noted that our study analyzed tracking data from a field experiment conducted by a single university team. This should be positioned as a first step toward understanding coordination processes related to mutual anticipation based on a common ground of team strategy.

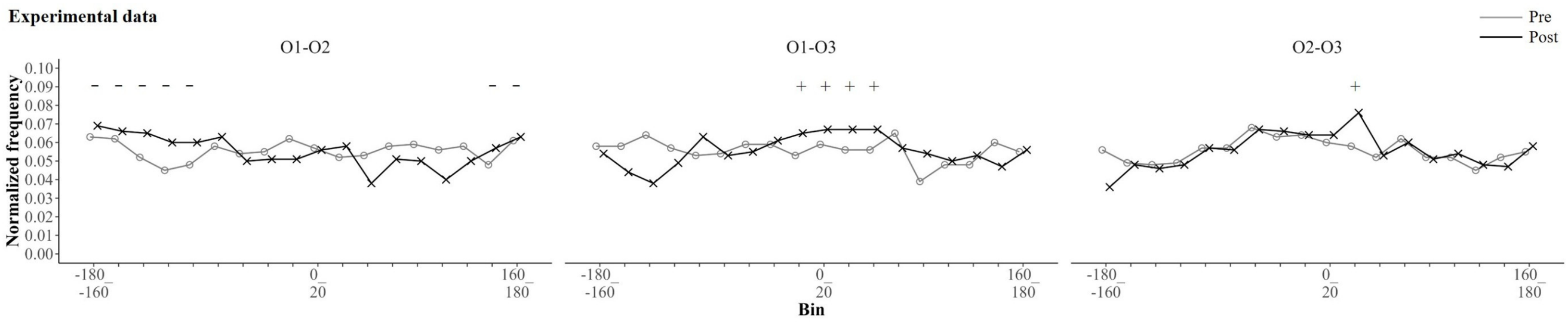

**FIGURE 4. Histograms representing the average frequency distributions of the relative phases across the game trials in the pretest and posttest conditions in the field experiment. "+" and "−" are indicated for the notable bins related to in-phase and anti-phase synchronization, respectively.**

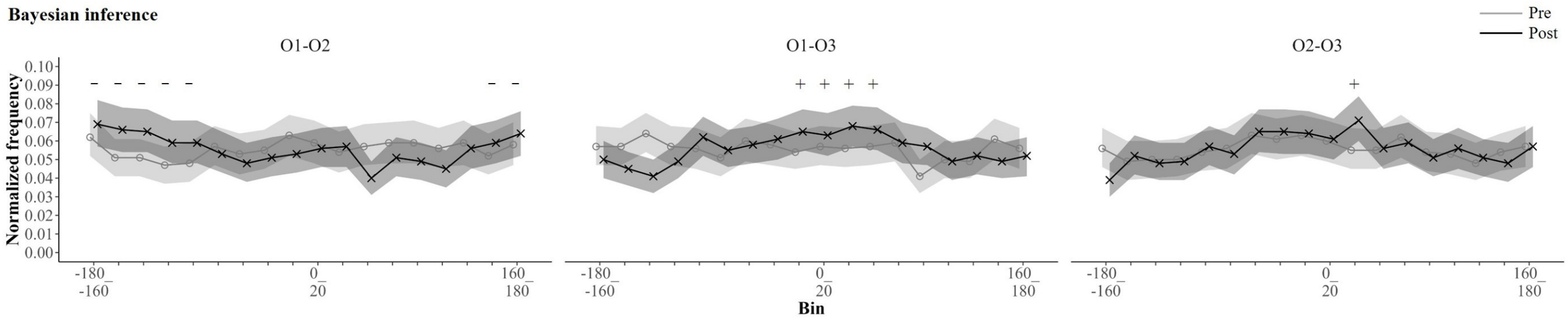

**FIGURE 5. Histograms representing the average frequency distributions of the relative phases across the game trials in the pretest and posttest conditions using Bayesian inference. Shaded bands represent 90% credible intervals. "+" and "−" are indicated for the notable bins related to in-phase and anti-phase synchronization, respectively.**

## V. CONCLUSION

The purpose of this cognitive science study was to extract the temporal components of movement in complex and dynamic team coordination involving heterogeneous roles and evolving interactions within short periods. We utilized a 3-on-3 basketball dataset from a study that examined offensive coordination, and calculated the relative phases of movements toward and away from each other. These frequency distributions were estimated using Bayesian inference, and the certainty of the direction of the difference before and after receiving advice from the coach was evaluated. The results showed that the posterior probabilities that the frequencies in synchronization-related bins were higher after the advice than before the advice were 70% or higher. These characteristics may be related to mutual anticipation based on a common ground of team strategy. However, this is only a possible interpretation in terms of cognitive science. Synchronization does not directly measure the degree of mutual anticipation. Future studies should conduct continuous measurements and analysis for teams from various sports and competitive levels. In addition, it is important to comprehensively investigate the relationships among synchronization, play events such as passing and dribbling, space structures created through player positioning, and team strategies. As a pilot study, the findings may provide a first step toward quantitatively understanding coordination processes.

## ACKNOWLEDGMENT

We would like to thank Masatoshi Yamada of Tokoha University for his valuable comments. The authors used ChatGPT to assist with English translation, grammar checking, and coding for statistical analysis. All outputs were reviewed and verified.

**TABLE 1. Bayesian inference of $m_{post} - m_{pre}$ and 90% credible intervals. For each pair of offensive players, these values are shown for each 20-degree bin in the pretest and posttest conditions.**

| O1–O2 | | | | O1–O3 | | | | O2–O3 | | | |
|---|---|---|---|---|---|---|---|---|---|---|---|
| Bin | $m_{post} - m_{pre}$ | 90% credible interval 5% | 95% | Bin | $m_{post} - m_{pre}$ | 90% credible interval 5% | 95% | Bin | $m_{post} - m_{pre}$ | 90% credible interval 5% | 95% |
| 1 | 0.007 | -0.010 | 0.024 | 1 | -0.008 | -0.022 | 0.007 | 1 | -0.017 | -0.032 | -0.003 |
| 2 | 0.015 | -0.001 | 0.032 | 2 | -0.012 | -0.026 | 0.001 | 2 | 0.002 | -0.012 | 0.016 |
| 3 | 0.015 | -0.001 | 0.031 | 3 | -0.024 | -0.037 | -0.010 | 3 | -0.001 | -0.015 | 0.013 |
| 4 | 0.012 | -0.002 | 0.027 | 4 | -0.008 | -0.022 | 0.006 | 4 | -0.001 | -0.015 | 0.013 |
| 5 | 0.011 | -0.004 | 0.027 | 5 | 0.007 | -0.008 | 0.021 | 5 | 0.003 | -0.012 | 0.018 |
| 6 | -0.004 | -0.020 | 0.012 | 6 | 0.004 | -0.010 | 0.018 | 6 | -0.003 | -0.018 | 0.011 |
| 7 | -0.005 | -0.019 | 0.009 | 7 | -0.002 | -0.017 | 0.013 | 7 | 0.002 | -0.014 | 0.017 |
| 8 | -0.004 | -0.018 | 0.011 | 8 | 0.003 | -0.013 | 0.017 | 8 | 0.004 | -0.012 | 0.020 |
| 9 | -0.010 | -0.026 | 0.006 | 9 | 0.011 | -0.004 | 0.026 | 9 | 0.002 | -0.015 | 0.018 |
| 10 | -0.003 | -0.018 | 0.013 | 10 | 0.006 | -0.009 | 0.021 | 10 | 0.001 | -0.014 | 0.017 |
| 11 | 0.003 | -0.013 | 0.019 | 11 | 0.012 | -0.003 | 0.027 | 11 | 0.016 | 0.000 | 0.032 |
| 12 | -0.017 | -0.032 | -0.003 | 12 | 0.009 | -0.006 | 0.024 | 12 | 0.001 | -0.015 | 0.016 |
| 13 | -0.008 | -0.023 | 0.007 | 13 | -0.001 | -0.016 | 0.015 | 13 | -0.003 | -0.019 | 0.013 |
| 14 | -0.011 | -0.026 | 0.005 | 14 | 0.016 | 0.003 | 0.029 | 14 | -0.003 | -0.018 | 0.012 |
| 15 | -0.012 | -0.026 | 0.003 | 15 | -0.001 | -0.015 | 0.013 | 15 | 0.003 | -0.012 | 0.018 |
| 16 | -0.003 | -0.018 | 0.012 | 16 | 0.003 | -0.011 | 0.017 | 16 | 0.002 | -0.012 | 0.016 |
| 17 | 0.006 | -0.009 | 0.022 | 17 | -0.012 | -0.026 | 0.002 | 17 | -0.006 | -0.021 | 0.008 |
| 18 | 0.006 | -0.010 | 0.022 | 18 | -0.004 | -0.018 | 0.010 | 18 | -0.001 | -0.017 | 0.015 |

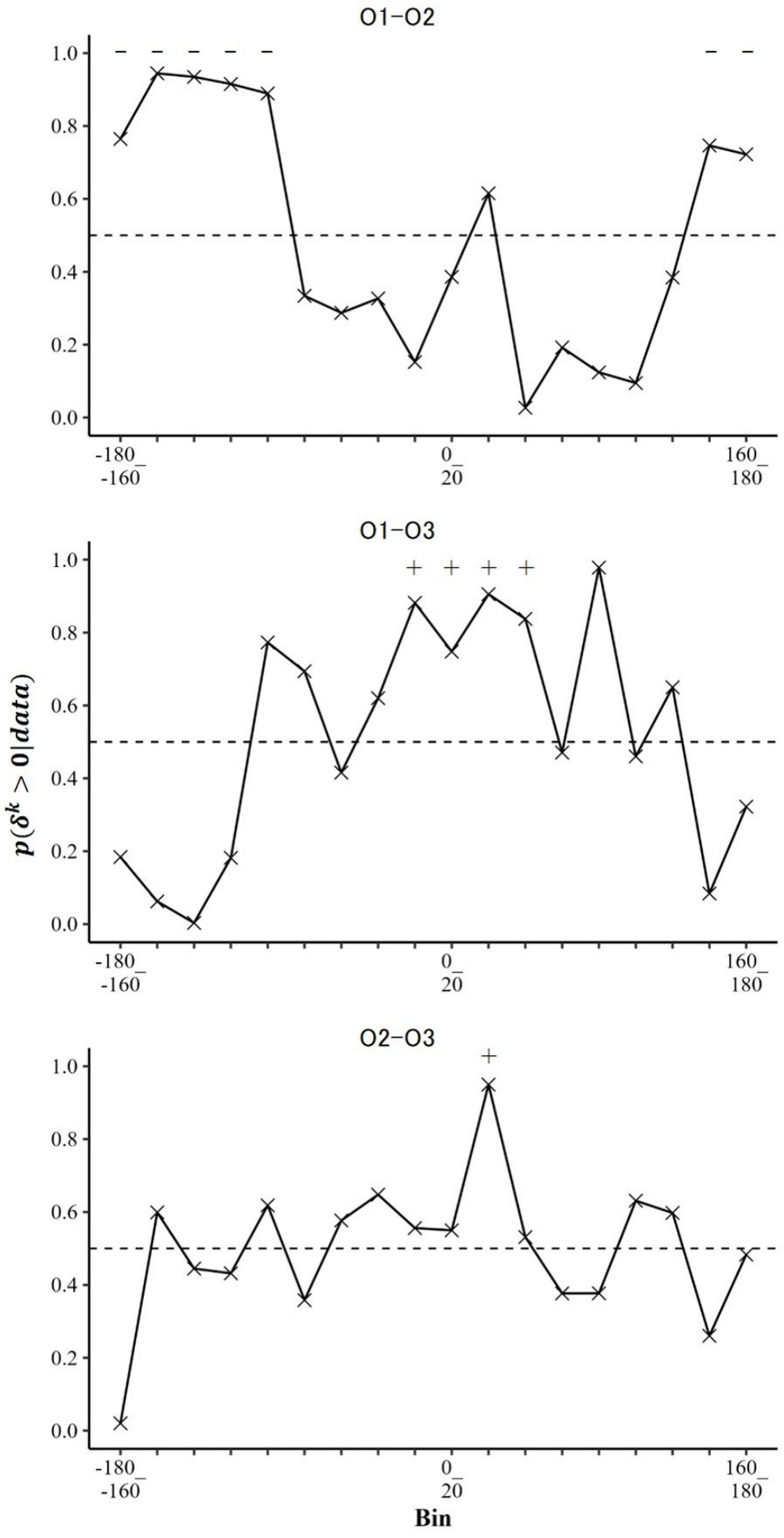

**FIGURE 6. Posterior probabilities that the frequencies in the posttest condition exceeded those in the pretest condition based on the observed data. The dotted line parallel to the horizontal axis indicates the chance level. "+" and "−" are indicated for the notable bins related to in-phase and anti-phase synchronization, respectively.**